\documentstyle[aps,prl,multicol]{revtex}

\newcommand{\beq}{\begin{eqnarray}}
\newcommand{\eeq}{\end{eqnarray}}
\newcommand{\n}{\nonumber}
\newcommand{\qo}{{(q,\omega)}}

\begin{document}

\title
{Sum rule analysis of Umklapp processes and Coulomb energy: \\
application to cuprate superconductivity}

\author{ Misha  Turlakov$^1$ and Anthony J. Leggett$^2$}
\address{$^1$Cavendish Laboratory, University of Cambridge, Cambridge, CB3 OHE, UK \\
$^2$Loomis Lab, University of Illinois at Urbana-Champaign, Urbana, Illinois 61801, USA}
\date{\today}
\maketitle

\begin{abstract}
The third moment frequency sum rule for the density-density correlation
function is rederived in the presence of Umklapp processes. 
Upper and lower bounds on the electron-electron Coulomb energy are derived
in  two-dimensional and three-dimensional media, and the Umklapp processes
are shown to be crucial in determining the spectrum of the density fluctuations
(especially for the two-dimensional systems).
This and other standard sum rules can be used in conjunction with experimental 
spectroscopies (electron-energy loss spectroscopy, optical ellipsometry, etc.)
to analyse changes of the electron-electron Coulomb energy
at the superconducting transition in cuprates.
\end{abstract}
\vskip2pc

\tighten

\begin{multicols}{2}
\section{Introduction.} 

Theoretical progress in the understanding of microscopic origin
of high-temperature superconductivity appears to be ambiguous.
The novelty and difficulty is to describe the strong effects
of the electron-electron interactions which determine  
the strongly correlated phases of cuprates at various dopings: 
the Mott antiferromagnetic insulator,
the ``anomalous'' metallic state and the superconductor.
In fact, the fundamentally new microscopic origin of superconductivity(SC)
in these materials is, perhaps, due to electron-electron interactions
unlike standard phonon-mediated superconductivity\cite{Schrieffer,remark}.
We explore the general aspects of such a scenario with the help
of sum rules for the density-density correlation function. 

The particular question of interest is the origin of the condensation energy,
the difference in the energy between the ``normal'' state extrapolated to $T=0~K$ and
the superconducting ground state. 
The famous BCS theory of superconductivity\cite{Schrieffer} is based on the 
assumption that the attractive interaction between electrons arises from
the lattice vibrations. Here we would like to investigate
a general alternative to the BCS phonon-mediated superconductivity, namely,
 that the superconducting state (e.g. the condensation energy)
is promoted either by the long-range part of electron-electron Coulomb interaction or by the 
short-range part of electron-electron  and static
electron-ion interactions.
The identification of the part of
the full electron-ion Hamiltonian responsible for the condensation energy would be an important
step towards a complete and consistent theory of  high-temperature
superconductivity. 

Many proposals for the condensation energy have been suggested\cite{Leggett,theories},
most of them (except Ref.\cite{Leggett}) being based on  phenomenological
Hamiltonians. Experimental confirmation of the origin of the condensation energy
 based on a phenomenological
Hamiltonian would still require the validation of the phenomenological Hamiltonian
(by various other experiments) in order to construct
a self-consistent theory.
Alternatively, the essential terms of 
the full\cite{comment-1} original electron-ion Hamiltonian
(and changes of the expectation values thereof upon transitions)
can be established consistently from experiments with the use of sum rules at the outset.
Such an approach, taken in this paper, can potentially identify
the origin of the condensation energy and the phenomenological Hamiltonian sufficiently
to describe  superconductivity and other strongly correlated phases.

If  the ion kinetic and ion-ion Coulomb energies are assumed irrelevant
(or in other words, these terms do not change upon the phase transitions
and do not determine the important correlated phases), 
the full\cite{comment-1} electron-ion Hamiltonian can be reduced to the following form

\beq
\hat{H}=\sum_{p,\sigma} \frac{p^2}{2m} c_{p,\sigma}^{+} c_{p,\sigma}
+\frac{1}{2\Omega} \sum_{q \neq 0} V_q [ \hat{\rho_q} \hat{\rho}_{-q}-N] + \n \\
+ \sum_{\kappa \neq 0} U_{-\kappa} \hat{\rho_\kappa},
\label{eq:hamiltonian}
\eeq
where $\hat{\rho_q}=\sum_{k,\sigma} c^{+}_{k-q,\sigma} c_{k,\sigma}=
\sum_{\vec{r_i}} e^{i\vec{q}\vec{r_i}}$ 
is the total density operator. $N$
is the number of electrons, and $\Omega$ is a total volume. 
The first and second terms are the kinetic and
Coulomb interaction energy of the electrons. The third term describes the interaction
of the electrons with
the periodic potential of the lattice, which can be
represented by the Umklapp pseudopotential $U_{-\kappa}$ 
with the sum over corresponding wavevectors $\kappa$ of the reciprocal lattice
\cite{quasimomentum}.
The term of interaction
between electrons and positive homogeneous ion background is omitted.
In spite of making
the ``static lattice'' assumption (the dynamic lattice effects (e.g. phonons)  are neglected),
 the Hamiltonian (\ref{eq:hamiltonian})
is quite general. For instance, the Hubbard model (and multi-band versions of it)
is only a reduced version of the Hamiltonian (\ref{eq:hamiltonian}) 
which neglects the long-range part of the Coulomb
interaction. Presumably, the Hamiltonian (\ref{eq:hamiltonian}) is sufficient
not only to describe the Mott insulating state and the metallic state at high densities
(far away from the half-filled band) but also all other important phases of the cuprates.

We can analyse the electron Coulomb energy 
in situations of different dimensionality. In an isotropic medium 
the three-dimensional (3D)
Coulomb potential is $V_q=\frac{e^2}{\epsilon_0 \epsilon_\infty q^2}$, where $\epsilon_\infty$
is the high-frequency dielectric constant due to the screening 
by  the core electrons\cite{epsilon}.
In a two-dimensional (2D) plane, the Coulomb potential is 
$V_q=\frac{e^2}{2\epsilon_0 \epsilon_\infty q}$.
For a layered electron gas
with  interplane distance $d$
(relevant for the discussion of single-layer  cuprates) 
the Coulomb interaction is

\beq
V_{q,q_z}= \frac{e^2 d}{2 \epsilon_0 \epsilon_\infty q} \frac{sinh(qd)}{cosh(qd)-cos(q_zd)} 
\label{eq:layered}
\eeq
with different dependencies on the wavenumbers $q$ parallel 
and $q_z$ perpendicular  to the planes\cite{Fetter}.
The Coulomb potential of a layered gas becomes three-dimensional
$V_q=\frac{e^2}{\epsilon_0 \epsilon_\infty q^2}$ in the long-wavelength limit
($q_z d \ll 1$ and $qd \ll 1$) and two-dimensional,
$V_q=\frac{e^2}{2\epsilon_0 \epsilon_\infty q}$, 
for short wavelengths ($qd \gg 1$, independently of $q_z$ momentum).

\section{Sum rules.} 
The electron-electron Coulomb energy can be related to the density
response function $\chi \qo$, and thus it is instructive to analyse various sum rules
for this response function\cite{assumption}. In particular,
the expectation value $<V_c>$ of the Coulomb energy can be written in the form
\beq
&&<V_c> \equiv \frac{1}{2\Omega} \sum_q V_q [<\hat{\rho_q} \hat{\rho_{-q}}>-N]= \n \\
&&=\sum_q [<V_{c,q}>-\frac{1}{2} V_q n], \n \\
&& <V_{c,q}>  \equiv \frac{1}{2} V_q  \int \frac{\hbar d\omega}{2\pi}  Im \chi \qo
coth \left( \frac{\hbar \omega}{2kT} \right),
\eeq
where $n=N/\Omega$ is the density of the electron system, and
the density-density correlation function $\chi \qo$\cite{comment-2}
 is defined in the standard way\cite{sign1,Nozieres}

\beq
\chi \qo \equiv \frac{i}{\hbar \Omega} \int_0^{+\infty} dt e^{i(\omega+i\delta)t}
<[\hat{\rho} (q,t),\hat{\rho}^{+}(q,0)]>. \label{eq:chi}
\eeq
$ <V_{c,q}>$ is the expectation value of the partial Coulomb energy 
corresponding to a particular value of momentum $q$.
The structure factor $S_q$

\beq
&&S_q=\frac{1}{N}<\hat{\rho_q} \hat{\rho_{-q}}>- N \delta_{q,0},
\eeq
can be expressed again through the imaginary part $Im \chi \qo$ using 
the fluctuation-dissipation theorem 

\beq
&&n S_q= \int_{-\infty}^{+\infty} 
Im \chi \qo coth \left( \frac{\hbar \omega}{2kT} \right) \frac{\hbar d\omega}{2\pi}, \n \\
&& <V_{c,q}>=\frac{1}{2} V_q n S_q.
\label{eq:formfactor}
\eeq

Various sum rules for the imaginary part of the susceptibility $Im \chi \qo$
(valid for arbitrary dimensional system with appropriate form of Coulomb potential $V_q$)
can be derived by calculating the commutators of the density operator and
the Hamiltonian.
To calculate the well-known $f$-sum rule (or the first moment sum rule)
it is necessary to calculate the expectation value of the operator
$[[\hat{\rho_q},\hat{H}],\hat{\rho_{-q}}]$. Another additional sum rule
is the causality (Kramers-Kronig) relation. These two well-known sum rules are
\cite{Nozieres}

\beq
&&J_{-1} \equiv \frac{2}{\pi} \int_0^{+\infty} \frac{Im \chi \qo}{\omega} d\omega= \chi (q,0), 
\label{eq:sum1} \\
&&J_1 \equiv \frac{2}{\pi} \int_0^{+\infty} \omega Im \chi \qo d\omega=
\frac{nq^2}{m}.
\label{eq:sum2}
\eeq
To calculate the $\omega^3$-moment sum rule\cite{Mihara}
the expectation value of the operator
$[[[[\hat{\rho_q},\hat{H}],\hat{H}],\hat{H}],\hat{\rho}_{-q}]$ must be calculated. 
The result of a long calculation  is

\beq
&&J_3 \equiv \frac{2}{\pi} \int_0^{+\infty} \omega^3 Im \chi \qo d\omega= \n \\ 
&&=\frac{1}{m^2} < \frac{1}{\Omega} \sum_\kappa (\vec{\kappa}\vec{q})^2 
(- U_{-\kappa} \hat{\rho_\kappa}) > + \n \\
&&+ q^4 \frac{n^2}{m^2} V_q +
q^4 \frac{3}{m^2} <\hat{T_{pr}}>  
+q^6 \frac{2n\hbar^2}{(2m)^3}+  \n \\
&&+ \frac{n}{m^2} \frac{1}{\Omega} \sum_{p \neq -q} [V_{|\vec{p}+\vec{q}|} (\vec{p}\vec{q}+q^2)^2-
V_p (\vec{p}\vec{q})^2 ] S_p,
\label{eq:sum3}
\eeq
where $\hat{T_{pr}}=\frac{1}{\Omega} \sum_p \frac{(\vec{p}\vec{\hat{q}})^2}{2m} c_p^{+} c_p=
\int \frac{d^3p}{(2\pi \hbar)^3} \frac{(\vec{p}\vec{\hat{q}})^2}{2m} c_p^{+} c_p $ 
is the projected kinetic energy operator,
 and $\vec{\hat{q}}=\vec{q}/|q|$ is a unit
vector along the direction of $\vec{q}$.
This higher order sum rule (Eqn.~\ref{eq:sum3}) is convenient for analysis
of high frequency transitions in the density response, because
it weights higher frequencies by a factor $\omega^3$.
The existence (or convergence) of the third moment sum rule can be demonstrated by showing
that all terms on the right-hand side of Eqn.~\ref{eq:sum3} are finite.
The first and the third terms are expected to be finite,
because they are essentially related to the finite expectation values
of the electron-lattice and the kinetic energy in the ground state.
The convergence and $q$-dependence of the last term will be discussed
in detail separately for 2D and 3D cases.



The three sum rules (Eqn.~\ref{eq:sum1}-\ref{eq:sum3}) allow us to derive
upper and lower bounds\cite{bounds} on the electron Coulomb energy,
and hence to discuss the possible changes of Coulomb energy due to
the phase transition.
Since the imaginary part of the susceptibility  $Im \chi \qo$ 
is a real positive definite function, the two Cauchy-Schwartz inequalities
can be written for the partial Coulomb energy $<V_c>_q$ at $T=0~K$

\beq 
\frac{1}{2} (V_q^2 J_{-1} J_1)^{1/2} \geq~~  <V_{c,q}>~~ \geq
\frac{1}{2} (V_q^2 \frac{J_1^3}{J_3})^{1/2}.
\label{eq:bounds}
\eeq
It is convenient to introduce the notional ``plasma frequency'' $\omega_p (q)$
defined by
$\omega_p (q)= \left( \frac{nq^2 V_q}{m} \right)^{1/2}$.
In fact, defined as above the ``plasma frequency'' has the right asymptotics
for the corresponding plasma waves
in the three-dimensional case ($\omega_{p,3D}^2 (q)=e^2 n_{3D}/(\epsilon_0 m)$) 
and the two-dimensional case
($\omega_{p,2D}^2 (q)= e^2 n_{2D} q/(2\epsilon_0 m)$). 
Another relation, which is useful in order to rewrite the $J_{-1}$ sum rule
for $q \rightarrow 0$, can be derived in the long-wavelength limit
($q \ll q_{TF}$, where $q_{TF}$ is the inverse of Thomas-Fermi screening length) 
for the full susceptibility $\chi (q,0)$, if we express
$\chi (q,\omega)$ through the ``bare''(or local) susceptibility $\chi_0 (q,\omega)$:

\beq
\chi (q,\omega)=\frac{\chi_0 (q,\omega)}{1+V_q \chi_0 (q,\omega)}
\label{eq:chi-sc}
\eeq
(this is an exact result, not an RPA approximation, provided $\chi_0$ is
defined in terms of the relevant irreducible diagrams).
The ``bare'' susceptibility $\chi_0 (q,0)$ of the electron liquid is assumed 
to be  finite  (see section 4.1 of Ref.\cite{Nozieres}  for the compressibility sum rule),
then for $q \ll q_{TF}$

\beq
\chi (q,\omega=0)=\frac{1}{V_q (1+\frac{1}{V_q \chi_0 (q,0)} )} \approx 
\frac{1}{V_q}.
\eeq 
The discussion up to this point is valid for any dimensionality of the system
(with the corresponding form of Coulomb potential $V_q$). In what follows
we analyse the two-dimensional and three-dimensional cases separately
and find important differences.

We consider first the 3D case.
The leading terms at small $q$ for the third moment
sum rule 
are

\beq
J_3 \approx \frac{q^2}{m^2} <\hat{A}>+
q^4 \frac{n^2}{m^2} V_q,
\label{eq:j3}
\eeq
where $\hat{A}=\frac{1}{\Omega} \sum_\kappa (\vec{\kappa}\vec{\hat{q}})^2  
(-1) U_{-\kappa} \hat{\rho_\kappa}$
\cite{sign}.
The third and  fourth terms in Eqn.~\ref{eq:sum3}, being proportional
to $q^4$ and $q^6$ powers, are subdominant.
The last term in (\ref{eq:sum3}) 
\beq
\frac{1}{m^2} \int d^D p [V_{|\vec{p}+\vec{q}|} (\vec{p}\vec{q}+q^2)^2-
V_p (\vec{p}\vec{q})^2 ] nS_p
\label{eq:last-term}
\eeq
requires careful analysis of its convergence and $q$-dependence.
The $q$-dependence (and convergence), due to the part of the integral over small momenta $p$,
is evident, since the upper bound on the pair-correlation
function $S_p$ (see Eqn.~\ref{eq:formfactor},~\ref{eq:bounds}) is
$S_p \leq Bp^{(D+1)/2}$ (where $B$ is a constant, $D$ is dimensionality)\cite{l-term}:

\beq
\int d^3 p V_p (\vec{p}\vec{q})^2 [S_{p+q}-S_p] \sim  \int d^3 p V_p (\vec{p}\vec{q})^2 S_q
\eeq
for small momenta $p<q$. Since in 3D case the upper bound $S_q \leq Bq^2$,
the contribution from small momenta $p$ integration  is at least of order $q^4$
(or higher power of $q$ for small $q$).
To analyse the $q$-dependence of the part of the integral (Eqn.~\ref{eq:last-term})
from  the integration
over large momenta $p$, we use the ``cusp theorem''\cite{Kaito},
which gives the asymptotic behaviour at large momenta $p$ (3D case)
\beq
S_p=1-\frac{C_{3D}}{p^4}+o(\frac{1}{p^5}),
\eeq 
where $C_{3D}$ is some constant. 
The integral of Eqn.~\ref{eq:last-term} with $S_p=1$
is identically zero (this is why $S_p$ can be substituted by $S_p-1$ if convenient).
It is easy to see that the integral over $p$ with the second term of expansion
$C_{3D}/p^4$ is convergent. By expanding in powers of $(q/p)$ for large $p$, 
the leading $q$-dependence of the considered term 

\beq
&&\int d^3 p V_p (\vec{p}\vec{q})^2 [S_{p+q}-S_p]
=\int d^3 p \frac{ V_p (\vec{p}\vec{q})^2 C_{3D}}{p^4} \ast \n \\
&& \ast \left( \frac{1}{(1+(q/p)^2+2(q/p)cos\theta)^2}   -1 \right)
\eeq 
is found to be proportional to $q^4$. The terms, proportional to $q^3$ and other odd
powers of $q$, are equal zero after the integration over angle $\theta$
($cos\theta=(\vec{p}\vec{q})/pq$ is the angle between
$\vec{p}$ and $\vec{q}$)). It can be seen that
for large $q$ the last term (Eqn.~\ref{eq:last-term}) grows no faster than $q^4$ as well.
What is important for the ensuing discussion is that, both for small and large $q$,
the last term of Eqn.~\ref{eq:sum3} has {\it subleading} 
$q^4$ dependence on the wavevector $q$.


The upper
and lower limits on the partial Coulomb energy can  be conveniently written
in terms of the 3D ``plasma frequency''

\beq
\frac{\hbar}{2} \omega_{p,3D} (q)+o(q^2) \geq ~ <V_{c,q}> , \n \\
<V_{c,q}> ~~ \geq ~
\frac{\hbar}{2} \frac{\omega_{p,3D} (q)}{\sqrt{1+\frac{<\hat{A}>}{n m \omega_{p,3D}^2 (q)}}}
+o(q^2).
\label{eq:ineq-3D}
\eeq
The inequalities (\ref{eq:ineq-3D}) have an interesting significance for
the possibility of gaining energy, in the SC condensation,
from small-$q$ modes. We first notice that in 3D case 
if $<\hat{A}>=0$ in both
normal and superconducting states, which is certainly the case if
there is no crystalline potential, 
the terms proportional to $q^4$ (the third and last terms of the right-hand side
of Eqn.~\ref{eq:sum3}) determine the difference between the upper and lower bounds.
Therefore if $<\hat{A}>=0$,
then for given $q$ the maximum
possible saving  
is proportional  to $(\hbar/2) \omega_{p,3D} (q/q_0)^2$      
(where $q_0 \sim q_{TF,3D}$, $q_{TF,3D}=\sqrt{\frac{e^2 k_F m^\star}{\epsilon_0 \pi^2 \hbar^2}}$ in 3D, 
$k_F$ is the Fermi wavevector), 
which is a negligible  portion of the partial Coulomb energy $\frac{\hbar}{2} \omega_{p,3D}$ 
at long wavelengths.
In fact, in the absence of the Umklapp processes the sum rules 
(Eqn.~\ref{eq:sum2} and ~\ref{eq:sum3}) essentially fix the density spectrum at long wavelengths
$q < q_{TF}$ 
to the plasma pole contribution: $Im \chi \qo \sim \frac{\pi \hbar \omega_{p,3D} (q)}{2 V_q}
\delta (\omega-\omega_{p,3D} (q))$, which satisfies completely all three sum rules 
at $q \rightarrow 0$ .
Other terms in the $J_3$ sum rule become comparable with the dominant term 
$q^4 V_q \frac{n^2}{m^2}$
only at $q \geq q_{TF}$. 
Therefore, a large saving of Coulomb energy {\it in 3D case 
in the absence of the lattice} due to the phase transition 
is possible  only for short wavelengths $q \geq q_{TF,3D}$. 


In the presence of the periodic lattice potential \\  ($<\hat{A}> \neq 0$), 
it can be seen from the inequalities (\ref{eq:ineq-3D})
that  for $q < q_{TF,3D}$
the maximum theoretical saving is a finite fraction of the zero-point 
plasma energy $\frac{\hbar}{2} \omega_{p,3D} (q)$.  
Therefore, {\it substantial saving of the Coulomb energy in the long-wavelength
limit is possible only in the presence of strong crystalline potential}.
In 3D (where the Coulomb interaction $V_q=e^2/\epsilon_0 q^2$), 
the Umklapp term $\frac{q^2}{m^2}<\hat{A}>$
contributes in the same leading order of powers of $q$ into the $J_3$ sum rule
as the Coulomb term $q^4 \frac{n^2}{m^2} V_q$ (see Eqn.~\ref{eq:j3}).
The Umklapp term is then responsible for the finite width (or ``lifetime'')
of the plasmon peak. 

{\it In the 2D case}, the leading terms are again

\beq
J_3= \frac{q^2}{m^2} <\hat{A}>+
q^4 \frac{n^2}{m^2} V_{q,2D}.
\eeq
Other terms are subleading and proportional to $q^4$ and $q^6$ powers.
The last term can be analysed similarly to 3D case and be shown proportional
to $q^4$ at small $q$.
Using the ``cusp theorem'' for 2D case\cite{Kaito}

\beq
S_p=1-\frac{C_{2D}}{p^3}+o(\frac{1}{p^4}),
\eeq 
we can expand in powers of $q/p$ for large momenta  $p$ and keep
the leading term as a function of $q/p$:

\beq
&&\int d^2 p V_{p,2D} (\vec{p}\vec{q})^2 
[S_{p+q}-S_p] \simeq 
\int d^2 p \frac{V_{p,2D} (\vec{p}\vec{q})^2 C_{2D} }{p^3} \ast \n \\
&& \ast \left( \frac{1}{(1+(q/p)^2+2(q/p)cos\theta)^{3/2} }-1 \right) .
\eeq
Due to the angle $\theta$ integration, terms proportional to $q^3$ and other odd
powers vanish. For large $q$, the last term grows no faster than $q^4$ also
(due to the first term in the bracket of Eqn.~\ref{eq:last-term}).
In the absence of Umklapp scattering ($<\hat{A}>=0$),
the density spectrum is given by the expression
$Im \chi \qo \sim \frac{\pi \hbar \omega_{p,2D} (q)}{2 V_{q,2D}}
\delta (\omega-\omega_{p,2D} (q))$, which satisfies all three sum rules
at small $q$. The maximum possible saving of partial Coulomb energy
at long wavelengths is of order $\frac{\hbar}{2} \omega_{p,2D} (q) \frac{q}{q_{TF,2D}}$
(where $q_{TF,2D}=\frac{e^2 m^\star}{2\pi \epsilon_0 \hbar^2}$ is a 2D Thomas-Fermi screening wavevector).

%
%

The presence of the Umklapp term $\frac{q^2}{m^2}<\hat{A}>$
has a much more dramatic effect on the density spectrum in 2D, 
because this term, proportional to $q^2$, has a leading power of $q$ in the third moment sum rule
at small $q$ dominating over the Coulomb term 
(unlike  3D case, where the Umklapp term has the same power-$q$
dependence as the Coulomb term). 
The density spectrum cannot be even approximated by a plasma pole expression,
and {\it the plasmon is  never really a well-defined excitation in 2D
in the presence of Umklapp scattering.}
It means that the Umklapp scattering modifies strongly (or ``non-perturbatively'')
the spectrum of the density fluctuations, and the spectrum is dominated
by multi-pair and pair excitations rather by plasmon.
Of course, the mere existence of a large value of $<\hat{A}>$ is in itself
perfectly
compatible with a traditional textbook picture, in which the sum rules are
satisfied by taking proper account of interband transitions; in such a case
there is no a priori reason why a plasmon associated with the excitations
of the conduction band must automatically be ill-defined. However, in a
system where Umklapp (quasi-momentum-nonconserving) interactions
between the Bloch quasiparticles are strong it seems natural that these
alone could give rise to a substantial value of $Im \chi$ even below the
first band gap; particularly in view of the above remark about the enhanced
effect in 2D, it is tempting to view the so-called midinfrared peak in the
cuprates in this light.

%
The upper bound on the partial
Coulomb energy is still the half of the plasmon energy ($\omega_{p,2D}(q) \sim \sqrt{q}$),
but the lower bound at small $q$ if $<\hat{A}> \neq 0$
is essentially given by

\beq
<V_{c,q}> ~ \geq~
\frac{\hbar}{2}  \frac{\omega_{p,2D} (q)}{\sqrt{1
+   \frac{<\hat{A}>}{n m \omega_{p,2D}^2 (q)}}}, 
\eeq
and so in the limit $q \rightarrow 0$,
\beq
<V_{c,q}> ~ \geq~ \frac{\hbar}{2} \omega_{p,2D}^2 (q) 
\left( \frac{nm}{<\hat{A}>} \right)^{1/2} \sim q.
\label{eq:2D-lowerbound}
\eeq
Therefore, bounds, based on sum rules, are compatible with saving of almost all Coulomb energy
$\frac{\hbar}{2} \omega_{p,2D} (q)$ in 2D when $<\hat{A}> \neq 0$.

It is necessary to mention extensive literature (for instance, Ref.\cite{Iwamoto})
using sum rules (in particular, third moment sum rule) in order to analyze
and derive various local-field corrections and approximations of the density response, whereas
our goal in this paper is to analyze general constraints on the electron
Coulomb energy at small $q$ without relying on any approximation.

It is also interesting to discuss briefly for comparison the interaction energy
of a many-particle system interacting via a short-range potential
($V_q \rightarrow V_0=const$, for $q \rightarrow 0$).
The upper bound on the partial interaction energy  $<V_{int,q}>$ is given by the ``acoustic mode''
$\frac{\hbar}{2} (V_0^2 \frac{nq^2}{m} \chi_0 (q,0))^{1/2}$

\beq 
\frac{\hbar}{2} \left( \frac{V_0 n}{m s} \right) q ~\geq~ <V_{int,q}>,
\eeq
where $\chi_0 (q,0)$ is given by the compressibility 
sum rule\cite{Nozieres}: $\chi_0 (q,0)=\frac{n}{ms^2}$,
where $s$ is the velocity of sound. Therefore, the maximum available
interaction energy at long wavelength is insignificant (especially when
weighted by the phase volume). For instance, it implies
that in most phase transitions in neutral systems
(i.e. many-particle systems interacting via a short-range potential)
the interaction energy is saved predominantly at short distances.





\section{Experimental probes.}
The remainder of this paper is devoted to a brief discussion of the experimental
spectroscopies which should, at least in principle, be able to shed light
on the origin of the condensation energy in the SC transition,
and the inferences which we may currently draw from them (cf. also Ref.\cite{Leggett},
section 4.2). For simplicity we will consider explicitly a single-plane
cuprate such as $Tl-2201$, so that in the normal phase there is only one
characteristic length (other than, possibly, the electron mean free path)
 large compared to the quantity $q^{-1}_{TF}$,
namely the interplane spacing $d$ (typically $\sim 10~\dot{A}$, i.e.
$\sim 10-20q^{-1}_{TF}$)(note notational differences from Ref.\cite{Leggett}).
The case of a multi-layer cuprate such as $Bi-2212$ is more complicated,
as there is now a second ``large'' characteristic length, namely
the intra-bilayer spacing ($\sim 3-5\dot{A}$); however, the general pattern
of the results is unchanged. In addition, we will assume tetragonal symmetry. 

The two spectroscopies which most directly probe the Coulomb energy,
or something closely related to it, are electron energy loss spectroscopy (EELS)\cite{Nucker}
and optical reflectivity\cite{raman}; in the latter case we shall assume
that ellipsometric measurements are possible in 
the interesting frequency regime\cite{Marel,Rubhausen} so that we may deduce 
the relevant complex dielectric constant without the use of Kramers-Kronig relations. 
In a bulk isotropic 3D metal the situation is very simple: to the approximation
that we neglect multiple scattering and the effect of the ionic cores,
the transmission EELS cross-section $\sigma(q,\omega)$ is a direct measure
of the quantity $V_q^2 Im \chi (q,\omega)$, where $\chi (q,\omega)$
is the ``true'' density susceptibility as defined in Eqn.~\ref{eq:chi}.
Since in 3D the longitudinal dielectric constant $\epsilon_{\|}(q,\omega)$
is identically equal to $1+\frac{e^2}{\epsilon_0 q^2} \chi_0 (q,\omega)$ with the help
of Eqn.~\ref{eq:chi-sc}, we find the simple result 

\beq
\sigma (q,\omega)=const \frac{1}{q^2} Im \left[ - \frac{1}{\epsilon_{\|}(q,\omega)} \right],
\label{eq:EELS}
\eeq
where the constant is of purely geometrical origin and can be calculated,
and the quantity $Im \left[ - \frac{1}{\epsilon_{\|}(q,\omega)} \right]$ is usually known
as the loss function. This formula is valid for arbitrary $q$, including
values larger the inverse lattice spacing in the approximation of neglect
of the direct scattering effect of the ionic cores.
If the latter is taken into account, the effect is to multiply the formula (\ref{eq:EELS})
by a factor which is in general a function of $q$ but not of $\omega$
provided that the latter is small compared to typical core excitation energies ($\sim 20eV$),
and expected to be unaffected by the superconducting transition.
Writing out the integrand of Eqn.~\ref{eq:formfactor} explicitly in terms
of $\chi_0  (q,\omega)$ (see Eqn.~\ref{eq:chi-sc}) and using the 3D relation
between the latter and $\epsilon_{\|}(q,\omega)$, we see that apart from a function
of $q$ {\it the transmission EELS cross-section is a direct measure of
the Coulomb energy locked up in $d\vec{q} d\omega$.}

It is well known that the condensation energy due to the SC transition is extremely small
(of order $10^{-4}~eV$ per electron (or per unit cell)) 
in comparison with the atomic energies ($10~eV$). It implies stringent requirements
on experimental techniques, nevertheless, changes associated with the SC transition
were observed by optics at mid-infrared frequencies\cite{Marel,Rubhausen} of magnitude
sufficient to provide the condensation energy 
(measured directly by specific heat measurements). It is worth noting that 
we do not discuss the changes
at frequencies comparable or lower than a superconducting gap
(although these changes of course are most remarkable consequences of superconductivity!),
because the change of Coulomb energy associated with this region of frequencies
is negligible (if limited to small momenta $q \ll q_{TF}$).
The optical reflectivity measurements\cite{Marel,Rubhausen}
have enough precision to explore the type of questions discussed 
in this paper, 
while it is hoped that the transmission EELS can achieve required accuracy
in the near future.

In a 3D bulk metal ellipsometric optical measurements can measure the complete
{\it transverse} dielectric constant $\epsilon_{\perp}(q,\omega)$, in the limit
$\vec{q} \rightarrow 0$, and hence the corresponding ``transverse'' loss
function $Im \left[ - \frac{1}{\epsilon_{\perp}(q,\omega)} \right]$.
Since in the normal phase, at least, there should be no distinction,
in the limit $\vec{q} \rightarrow 0$, between $\epsilon_{\|}(q,\omega)$ and
$\epsilon_{\perp}(q,\omega)$, it follows that in this phase
the $\vec{q} \rightarrow 0$ limit of ``loss functions'' measured by EELS and by optics
should coincide. It is a somewhat delicate question, once one renounces reliance on some
specific model such as the Fermi liquid one, what is to count as 
``the  $\vec{q} \rightarrow 0$ limit''; in addition to the obvious scale
$q_{TF}$ or $q_F$, it is not immediately clear that the inverse electron mean free path
$1/l$ might not be a relevant quantity. However, it is plausible that
this quantity should not play a major role for $\omega$ in the mid-infrared region,
so that we shall tentatively take ``$\vec{q} \rightarrow 0$'' to mean
in the 3D case $q \ll q_{TF},q_F$.

Some care is needed in adapting the above results to the case of a layered material
such as cuprates, even if we specialize (as we shall) to the limit
$q_z \ll 1/d$, which is automatically fulfilled in optical experiments and
may be satisfied in transmission EELS by a suitable choice of geometry.
In the context of EELS experiments, we now have to distinguish the cases
$qd \ll 1$ and $qd \gg 1$ (where $q$ is the ab-plane component of the momentum loss).
In the former case the 3D bulk formula (Eqn.~\ref{eq:EELS}) applies
unchanged, provided that $\epsilon_{\|}(q,\omega)$ is defined to be
the tensor component of the longitudinal dielectric constant
corresponding to current flow in the ab-plane; note in particular that,
at least in the normal phase, we expect that  
$Im \left[ - \frac{1}{\epsilon_{\|}(q,\omega)} \right]$ is nearly independent
of $|\vec{q}|$ in the limit $\vec{q} \rightarrow 0$.
In the opposite limit $qd \gg 1$, we could choose 
to continue to use the 3D formula(Eqn.~\ref{eq:EELS}), but would
than find that the $\epsilon_{\|}(q,\omega)$ so defined has a strong explicit
dependence on $q$. A much more natural convention in this limit is to
treat the scattering as occuring independently from the different $CuO_2$ planes,
and to define a {\it two-dimensional} (``per-plane'') version
$\chi_0^{(2)} (q,\omega)$ of  $\chi_0 (q,\omega)$, or equivalently
a quantity (cf. Ref.\cite{Leggett}), section 4.1)

\beq
K(q,\omega) \equiv \frac{1}{2\epsilon_0 q^2} \chi_0^{(2)} (q,\omega)=\frac{d}{2}
(\epsilon_{\|}(q,\omega)-\epsilon_b),
\eeq
where $\epsilon_{\|}(q,\omega)$ is the ``natural'' definition of the 3D bulk
ab-plane dielectric constant, i.e. the quantity which relates
the local polarization to the local field, and is expected to be nearly
constant over a range $q \gg 1/d$, and $\epsilon_b$ is the ``background'' (off-plane)
contribution to it (cf. Ref.\cite{Leggett}).
With this definition we find that, apart from factors depending
only on $q$, both the transmission EELS cross-section and the (single-plane)
Coulomb energy locked up in the range $d\vec{q} d\omega$ are proportional to the quantity

\beq
-Im \left( \frac{1}{1+qK(q,\omega)/\epsilon_{sc}} \right),
\label{eq:2D-EELS}
\eeq 
where $\epsilon_{sc}$ is the dielectric constant which screens the Coulomb
interactions of the in-plane electrons (note that in general $\epsilon_{sc}$
is not equal to $\epsilon_b$). Thus, just as in the bulk 3D case, the transmission EELS
cross-section is a direct measure of the Coulomb energy locked up
in the relevant region of $(q,\omega)$-space; note however that Eqn.~\ref{eq:2D-EELS}
introduces an extra explicit $q$-dependence which absent
in the bulk case. This subtlety seems to have been overlooked in the analysis
of existing normal state EELS data\cite{Nucker} on the cuprates,
where it seems to be assumed that even in the regime $qd \gg 1$ EELS experiments
measure the ``bulk''  $\epsilon_{\|}(q,\omega)$. In the intermediate case ($qd \sim 1$)
a similar analysis using Eqn.~\ref{eq:layered} is possible, but will not be given here.

In the case of optical experiments on the cuprates (with the sample surface assumed
to lie in the ab-plane) we always have $qd \ll 1$, $q_z d \ll 1$ , and
thus at first sight we would expect the complex dielectric constant inferred from
reflectivity measurements to be identical to the $\epsilon_{\|}(q,\omega)$
inferred, via Eqn.~\ref{eq:EELS},\ref{eq:2D-EELS}, from EELS experiments in the limit
$\vec{q} \rightarrow 0$. Existing measurements in the normal phase do
appear to be consistent with this prediction. In the superconducting state, however,
there are three complications: first, it is not clear that even in the mid-infrared
regime the Cooper-pair radius $\xi_0$ is not a relevant length scale, so that it may be
illegitimate to use the ``true'' $\vec{q} \rightarrow 0$ ($q \xi_0 \ll 0$) behaviour
observed in the optics to infer the behaviour in the regime
$1/\xi_0 \ll q \ll q_{TF}$. Secondly,
it is not completely obvious, particularly in the former limit $q \xi_0 \ll 1$,
that the finite-frequency longitudinal and transverse dielectric constants
must be equal in the superconducting state. This latter complication
may be somewhat mitigated by a third consideration, namely that at the non-normal
angles of incidence necessary in the ellipsometric technique what is measured,
 in a layered material, is not simply $\epsilon_{\perp}$ but a combination
of $\epsilon_{\perp}$ and $\epsilon_{\|}$!
We will not attempt to develop those points further here, but will
rather use them to draw the conclusion that, while the spectacular changes
observed\cite{Marel,Rubhausen} in the optically measured dielectric constants
of the cuprates at and below the superconducting transition are strongly suggestive,
a quantitative test of any scenario (such as the mid-infrared one of Ref.~\cite{Leggett})
which attributes the energy saving largely to a regime of $q$ small
compared to $q_{TF}$ but large compared to $\xi_0$ (and $1/d$) will require
accurate transmission EELS data taken across the transition, something which
(as regards the mid-infrared regime of frequencies) does not to our knowledge
at present exist.

We finally address head-on the question: which of the three terms in Eqn.~\ref{eq:hamiltonian}
is (are) reduced in the superconducting transition? If it is the second (Coulomb) one, 
in what regime(s) of $q$ and $\omega$ does the saving predominantly occur? Part of the interest
of this question is that as we have seen above, a conjectured answer can be tested directly
in transmission EELS experiments.

We start by recalling a well-known result: since the original (``true'') Hamiltonian
of the N-body system (the nonrelativistic limit of the Dirac Hamiltonian) is composed exclusively
of kinetic energy terms and (unscreened) Coulomb interactions, the virial theorem
immediately tells us that the change in total kinetic energy (of electrons
and ions) must be exactly minus half that of the total Coulomb energy 
(electron-electron, electron-ion and ion-ion), and thus Coulomb energy must be saved
in the superconducting transition (and indeed in any other phase transition
into a lower-energy state). While this conclusion is very generic and rigorous,
it is not usually regarded as giving much insight into the ``mechanism''
of superconductivity in the cuprates (or for that matter in the classic superconductors)
since the term ``mechanism'' is often held to refer to a low-energy
effective Hamiltonian in which the separation of the original
kinetic and potential energies may no longer be explicit.
However, ``intermediate-level'' effective Hamiltonian (\ref{eq:hamiltonian})
is sufficiently close to the original truly first-principles one that the virial-theorem result
for the latter might at least suggest that it is one or both of the last two
terms of  (\ref{eq:hamiltonian}) which are saved\cite{ambiguity}.


If we assume for the sake of argument that Coulomb energy is indeed saved, 
then where in the space of $q$ and $\omega$ is it saved? It is at this point
that the sum rules arguments of section 2 come into their own.
For convenience we reproduce here the three relevant sum rules with terms 
of relative order $q^4$ and higher omitted on the right-hand sides: with notation
as in Eqns.~\ref{eq:sum1},\ref{eq:sum2},\ref{eq:sum3};

\beq
&&J_{-1}=\frac{1}{V_q}, \label{eq:tr-j1} \\
&&J_1=\frac{nq^2}{m},   \label{eq:tr-j2} \\
&&J_3=\frac{n^2 q^4}{m^2} V_q + \frac{q^2}{m^2} <\hat{A}>  \label{eq:tr-j3},
\eeq
where in the case of a layered system $V_q$ is given by Eqn.~\ref{eq:layered},
and tends to $e^2/(\epsilon_0 \epsilon_{\infty} q^2)$ for $qd \ll 1$ (and $q_z d \ll 1$, see above)
and to $e^2/(2 \epsilon_0 \epsilon_{\infty} q)$ for $qd \gg 1$, and where $\hat{A}$ is defined below
Eqn.~\ref{eq:j3}. The ($T=0~K$) contribution $<V_{c,q}>$ to the expectation value
of the Coulomb energy from wave vector $\vec{q}$ is, up to a factor, just $J_0$
(see Eqn.~\ref{eq:formfactor}).

The arguments of section 2 now show that the maximum change of $<V_{c,q}>$
in the ``essentially 3D'' regime  $qd \ll 1$ is proportional to a finite
fraction of the ``3D plasmon energy'' but weighted by $q^2$ of the phase space ($d^3 q$),
and hence the maximum total saving possible from this regime is very small.
On the other hand, the contribution from the regime $1/d \ll q \ll q_{TF}$
(where the truncated forms Eqns.~\ref{eq:tr-j1}-\ref{eq:tr-j3} are still a good
approximation) can be of order $q^{1/2}$ (cf. the conclusion after Eqn.~\ref{eq:2D-lowerbound}),
 provided the quantity $<\hat{A}>$ is substantial, while the phase space
allows significant saving of Coulomb energy since $dq_z d^2 q \sim (2\pi/d) 2\pi qdq$ and
 $1/d \ll q \ll q_{TF}$.
Thus, in a quasi-2D system with a large value of  $<\hat{A}>$ substantial energy
is in principle available for saving in this small-$q$ regime. To estimate
the value of $<\hat{A}>$ we return for a moment to the limit  $qd \ll 1$
and refer to the normal state optical loss function data: 
using Eqns.~\ref{eq:tr-j1}-\ref{eq:tr-j3}
with the appropriate form $(e^2/(\epsilon_0 \epsilon_{\infty} q^2))$ of $V_q$,
we see that the quantity $<\hat{A}>_n$ is given, in the natural units of 
$n^2 e^2/(\epsilon_0 \epsilon_{\infty})$ by the expression (``n'' is  the normal-state value)

\beq
\frac{<\hat{A}>_n}{n^2 e^2/(\epsilon_0 \epsilon_{\infty})} = J_3-J_1^2/J_{-1}.
\label{eq:An}
\eeq
Although a strict evaluation of the right-hand side of Eqn.~\ref{eq:An}
from the optical loss-function requires us to know about the effective
frequency cutoff (since at high frequencies there will be contributions to
$\epsilon (q,\omega)$ from ``core'' processes not described by Eqn.~\ref{eq:hamiltonian}),
it is clear that the mere existence of  mid-infrared(MIR) peak extending over 
an order of magnitude in frequency already implies that it at least of order of 1. Thus,
a very appreciable fraction of the Coulomb energy locked up, in the normal state,
in the low-$q$, MIR-frequency regime is in principle available for saving in the
SC transition(or indeed in other possible phase transitions). Whether
it is in fact saved and to what extent, as in fact postulated in the ``MIR scenario''
of Ref.\cite{Leggett}, depends of course on the cost of the formation of the
 Cooper pairs in kinetic and/or static lattice energy.
Actually, rather than asking as above for the fraction of the Coulomb
energy which is in principle available for saving (something which is
not that significant if the original value is itself small), it may be more
informative to estimate the relative contribution of the small-$q$ regime
in 2D and 3D {\it for a given change in $\chi_0$} due to the phase transition.
Taking into account both the phase space factor and the extra factor
of $q$ in the denominator of the expression (\ref{eq:2D-EELS}) in the 2D case, 
we find that in the regime
where the $(qK)$ dominates the contribution of small $q$ is proportional to
$q^2$ in 3D but to a constant in 2D, so that the relative importance of the
long-wavelength regime is vastly enhanced in the 2D case.

On the experimental front, it has to be said that as noted in Ref.\cite{Marel},
the optical data, if extrapolated into the relevant ($q \xi_0 \gg 1$) regime
with several other assumptions, indicate rather the opposite, i.e. that
the Coulomb energy associated with the MIR regime actually increases
in the SC state. However, because of the various considerations noted above,
this extrapolation may be problematic, and a definitive test of 
the MIR hypothesis must await quantitative transmission EELS measurements
across the superconducting transition (or a better theoretical
understanding of the generic $q$-dependence of $\epsilon (q,\omega)$
in the SC state).

In sum, we analyzed the electron-electron Coulomb energy in the presence
of the periodic lattice potential using various sum rules for the density-density
response function. We believe that in this paper we have made it plausible
that two specific properties of cuprates, namely, (a) the layered (two-dimensional)
structure of the $CuO_2$ planes and (b) the occurrence of a broad and strong
peak in the optical loss function, can be essential ingredients
in the occurrence of high-temperature superconductivity in these materials
by conspiring to the saving of small-$q$ ($q< q_{TF}$) part of 
electron-electron Coulomb energy.

We are grateful for discussions to Dirk van der Marel and Neil Ashcroft.
A.J.L. acknowledges the support of grant NSF-DMR-99-86199.
M.T. was supported by EPSRC.

\end{multicols}

\begin{references}
\bibitem{remark}
Although the phonon mechanism for   superconductivity
is not ruled out irrevocably for the cuprate superconductors, we explore the general
scenario of an electronic mechanism, at least, as a matter of principle.
\bibitem{comment-1}
The separation into conduction and core electrons is assumed,
where the core electrons contribute to the screening constant $\epsilon_\infty$ 
and to the effective periodic potential.
For simplicity, we do not consider here possible effects of 
anisotropic (or $q$-dependent) screening by the core electrons. 
We also neglect the effects of disorder.
\bibitem{quasimomentum}
Throughout this paper we do not use the reduced zone scheme (in order not to be based
on this one-electron approximation in the presence of strong el-el interaction),
and all momenta are real (not quasimomenta).
Here the terms, Umklapp and lattice scattering, are used interchangeably
to denote the static effects of the lattice potential.
In the reduced zone scheme,
one can distinguish between ``lattice'' and ``Umklapp'' effects,
the former  being the  real momentum-non-conserving effects of the static lattice,
while the latter due to effects of el-el Coulomb interaction
(rigorously conserving real momentum, but in general not conserving quasimomentum).
\bibitem{epsilon}
In the analysis of upper and lower bounds on the electron Coulomb energy 
in the text of the paper we omit the
constant $\epsilon_\infty$ in order to keep expressions simple.
The modifications due to the dielectric constant $\epsilon_\infty$  are trivial.
\bibitem{assumption}
The sum rule analysis is applicable only to homogeneous phases of cuprates.
\bibitem{comment-2}
Of course, in a system with crystalline periodicity the coordinate-space
form of $\chi$ is a function of the center-of-mass coordinate as well as the
relative one, and the momentum-space form is correspondingly a matrix
quantity $\chi(q+\kappa_i,q+\kappa_j)$, where $\kappa_{i,j}$ 
are reciprocal lattice vectors. The quantity
discussed everywhere in the text  is in this notation $\chi(q,q)$.
\bibitem{sign1}
We take here a definition of the density correlation function such that
for positive frequencies ($\omega > 0$) the imaginary part of the susceptibility
is positive $Im \chi (q,\omega) > 0$, and also $Re \chi (q, 0) >0$.
This definition is different by a minus sign from the definition in Ref.\cite{Nozieres}
(section 4.1). See also below, that $\chi_0$, defined in Eqn.~\ref{eq:chi-sc}, is different 
only by minus sign from $\chi_{sc}$ defined in Ref.\cite{Nozieres} (section 4.1).
\bibitem{sign}
Here we give arguments that the expectation value of $<\hat{A}>$
is positive. Namely, if $U(x) \sim U_0 cos(\kappa x)$, the electron density 
minimises the interaction energy by taking the form
$\rho(x) \sim - \rho_0 cos(\kappa x)$. In other words, the electron-lattice
energy $<\sum_\kappa U_{-\kappa} \hat{\rho_\kappa} >$ (without 
the zero-momentum ``background'' harmonic $\kappa=0$) is always negative,
because the harmonics of the density $\rho_\kappa$ 
are shifted by the phase $\pi$ to the harmonics of the lattice potential $U_{-\kappa}$.
In addition, from inequalities Eqn.~\ref{eq:bounds},\ref{eq:ineq-3D},\ref{eq:2D-lowerbound} 
it follows that
the expectation value $<\hat{A}>$ is positive.
\bibitem{l-term}
It can be convenient for the analysis to rewrite Eqn.~\ref{eq:last-term} as
$(n/m^2) \int d^D p V_p (\vec{p}\vec{q})^2  [S_{p+q} -S_p]$ by making
a change of variable $p'=p+q$ in the first term in the bracket of Eqn.~\ref{eq:last-term}.
\bibitem{raman}
Raman scattering is also relevant, but is a less direct probe
and will not be discussed here.
\bibitem{ambiguity}
Of course we cannot apply the virial theorem directly to Hamiltonian (Eqn.~\ref{eq:hamiltonian}),
because the lattice potential $U_\kappa$ is not in general a sum of simple
Coulomb terms.
\end{references}
\end{document}